\documentclass[11pt,a4paper]{article}
\usepackage{jinstpub}
\usepackage[utf8]{inputenc}
\usepackage{subfig}

\title{Double cascade reconstruction in KM3NeT/ARCA}

\author[a,1]{T.~van Eeden%
\note{Corresponding author.}}
\author[a]{and A.~Heijboer}
\affiliation[a]{Nikhef,\\
Science Park 105, 1098 XG Amsterdam, Netherlands}

\emailAdd{tjuanve@nikhef.nl}

\collaboration[c]{on behalf of the KM3NeT collaboration}

\proceeding{VLVnT 2021 - Very Large Volume Neutrino Telescope Workshop 2021\\
  18 - 21 May 2021\\
  Valencia}

\abstract{
The detection of astrophysical $\nu_\tau$ is an important verification of the observed flux of high-energy neutrinos. A flavour ratio of approximately $\nu_{e} : \nu_\mu : \nu_\tau \approx 1 : 1 : 1$ is predicted for astrophysical neutrinos measured at Earth due to neutrino oscillations. On top of this, the $\nu_\tau$ offers a unique channel for neutrino astronomy due to absence of an atmospheric $\nu_\tau$ background contribution. When a $\nu_\tau$ interacts it produces a particle cascade and often a $\tau$ lepton which in turn decays mainly into another cascade. This results in a double cascade signature. An excellent angular resolution can be achieved when both cascade vertices are reconstructed. The KM3NeT/ARCA detector, which is under construction in the Mediterranean sea, will be able to detect this signature due to its timing and spatial resolution for cascades. We will discuss the dedicated reconstruction algorithm and performance for reconstructing double cascades using KM3NeT. The angular deviation reaches sub-degree level for tau lengths larger than 25 meters.}


\begin{document}

\maketitle

\section{Introduction}

The KM3NeT/ARCA detector is currently under construction at the bottom of the Mediterrean sea near Portopalo di Capo Passero on Sicily, Italy \cite{loi}. The detector consists of a 3-D grid of optical modules that each contain 31 photomultiplier tubes (PMTs). The timing resolution offers opportunities in the identification and reconstruction of tau neutrino interactions. Tau neutrinos can interact through the charged current (CC) weak interaction where it will produce a tau lepton and a particle cascade from the shattered nucleon. The tau has a mean lifetime of $2.903 \pm 0.005 \times 10^{-13}$ seconds and it decays into hadrons and leptons \cite{taupartdata}. The branching ratio to an electron or hadrons is 0.8261 and this results in a double cascade signature. The cascades are separated by an average $5 \frac{ \text{cm}}{\text{TeV}}$ due to time dilatation. In this work, we present a new reconstruction algorithm for double cascades. We find an improved angular reconstruction performance with respect to single cascades due to the \textit{lever-arm effect}. The early and late part of the event are strongly restricted in position thanks to the arrival time of the light that they produce. This places the start and end of the event along the direction of the tau lepton, resulting in a better angular resolution.

\section{Event simulation}

The double cascades from $\nu_\tau$ charged current (CC) interactions were simulated using gSeaGen. The gSeaGen code is a GENIE-based application developed to generate events induced by neutrino interactions \cite{gseagen}. The events were subsequently processed using internal KM3NeT software for Cherenkov light generation in seawater, photomultiplier tube (PMT) simulation and triggering.

Events were selected with criteria based on the standard KM3NeT single cascade reconstruction: Aashowerfit. The reconstructed vertex is required to be inside the instrumented volume of the detector and the reconstructed energy above 100 TeV.

\section{Method}

The double cascade reconstruction algorithm consists of three stages as described below.

\subsection{Single cascade reconstruction}

The Aashowerfit algorithm utilises information on which PMTs where hit and not hit due to a cascade event. It subsequently fits the spatial Cherenkov profile of light to the data to get an estimate for the direction and the energy of the neutrino. The arrival time of light information is used to get an estimate for the vertex position and time. The Aashowerfit algorithm assumes a single cascade topology which is used as a direction, energy and vertex prefit for double cascade topologies.

\subsection{Tau length prefit}

The tau length prefit consists of a likelihood scan along the direction fitted by Aashowerfit. The algorithm assumes two cascades with equal energy at the Aashowerfit vertex and starts varying the position and time of both cascades. The positions and times are constrained by the Aashowerfit direction and the speed of light because the tau lepton is highly relativistic. The algorithm maximises the likelihood
\begin{equation}
    \text{Likelihood} = \prod_{\text{all hits}} 1 - e^{-n(t) - R_{bg}}
\end{equation}
where $R_{bg}$ is the background rate and $n(t)$ is the expect photon density at a given time $t$ that is obtained from tabulated probability density functions (PDFs). All hits are selected within a cylinder with a radius of 300 meters surrounding the Aashowerfit direction. The length of the cylinder extends to the edges of the instrumented volume.  The reconstruction is a maximum likelihood estimator of the tau propagation length and it provides two offsets from the Aashowerfit vertex along the Aashowerfit direction. This results in an estimate for the neutrino interaction vertex and the tau decay vertex.

\subsection{Double cascade full fit}

The double cascade full fit adopts the starting values from the previous steps and performs a likelihood fit where the following parameters are free:
\begin{itemize}
    \item Neutrino interaction vertex (x,y,z,t)
    \item Direction ($\theta,\phi$)
    \item Tau length (len)
    \item Energy asymmetry ($\frac{\text{E}_1 - \text{E}_2}{ \text{E}_1 + \text{E}_2 }$).
\end{itemize}
The algorithm assumes two colinear cascades correlated by the speed of light due to the highly-relativistic tau lepton. Aashowerfit provides the estimation for the total energy and the double cascade fit finds the energy division between both cascades through the energy asymmetry. The likelihood that is maximised is defined as
\begin{equation}
    \text{Likelihood} = \prod_{\text{1st hits}} \text{P}_{ \text{1st} } \text{ (t)}
\end{equation}
where $\text{P}_{ \text{1st} }$ is the probability density for the first hit to occur at time $t$ given that a hit occurs. The likelihood is calculated using the first hits on every PMT starting from -20 ns with respect to the fitted time of the neutrino interaction vertex of the tau length prefit. We select the first hits because they do not contain timing effects of the signal processing. In KM3NeT, the analogue pulses from the PMTs are digitised into a \textit{hit arrival time} and \textit{time-over-threshold} (ToT) \cite{frontend}. Consecutive signals on a PMT can be merged into a hit with a larger ToT complicating the use of all hits. The double cascade full fit therefore uses the first hits in order to utilise reliable timing information. This is not yet possible for the tau length prefit because the use of first hit information requires a reliable estimation of the starting time of an event.

\section{Performance}

\begin{figure}[h]
\centering
\includegraphics[width=0.7\textwidth]{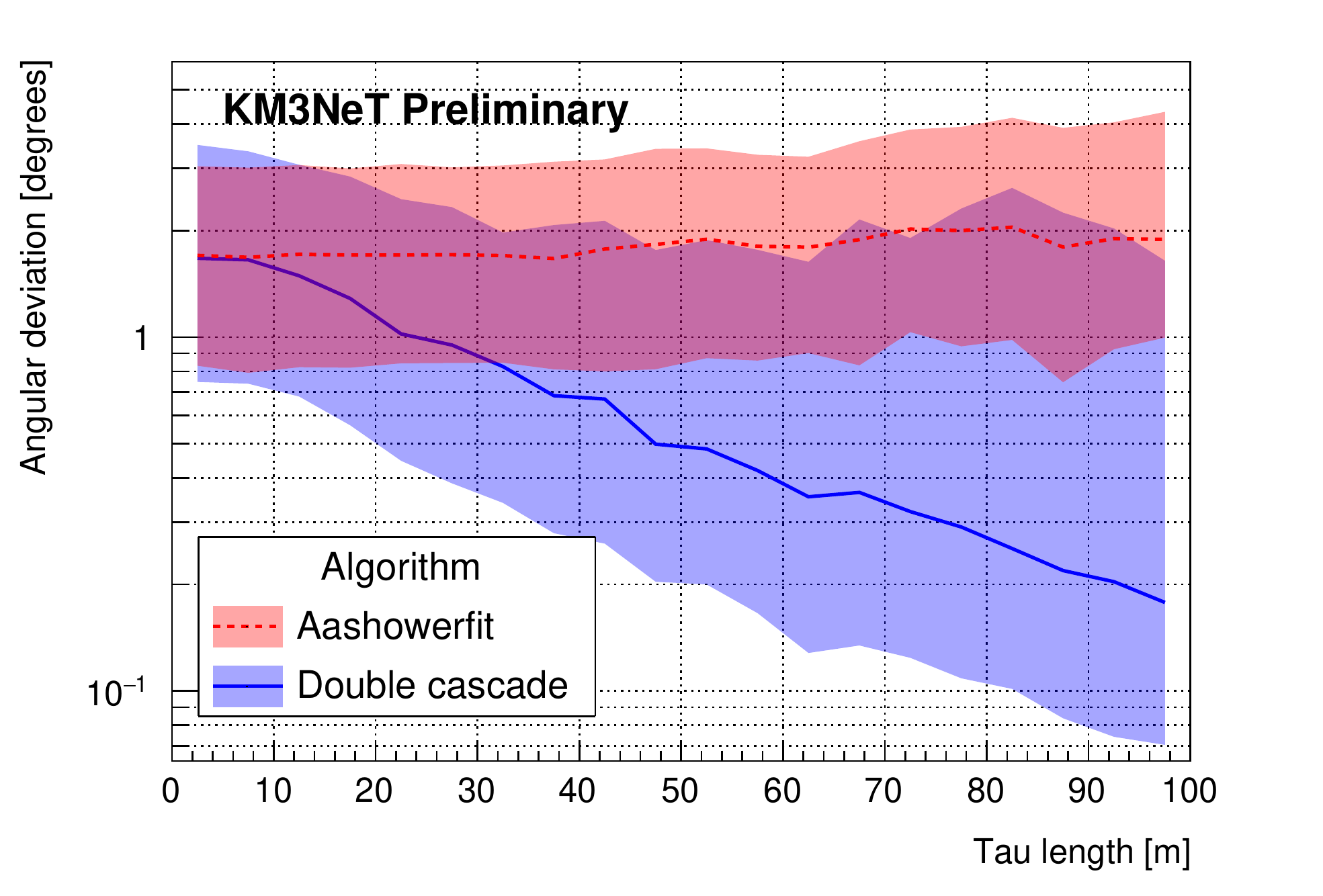}
\caption{Median and 68\% quantiles of the angular deviation for Aashowerfit and the double cascade reconstruction for selected double cascade events.}
\label{fig:angres}
\end{figure}

Figure \ref{fig:angres} shows the angular deviation of the double cascade reconstruction algorithm on the selected double cascade events. The angular deviation is defined as the angle between the reconstructed direction and the true neutrino direction. The double cascade performance is compared with the Aashowerfit performance for the same events to show the merit of reconstructing double cascade events with a double cascade hypothesis. The median of Aashowerfit stays at $\sim$ 2 degrees for all tau lengths, while the double cascade reconstruction drops below 1 degree for tau lengths larger than 25 meters and reaches 0.2 degrees for tau lengths of 100 meters.

\begin{figure}[h]%
    \centering
    \subfloat[\centering Reconstructed length error for the double cascade reconstruction. The median and 68\% quantiles are $0.72^{+1.23}_{-1.95}$ meters.]{{\includegraphics[width=7cm]{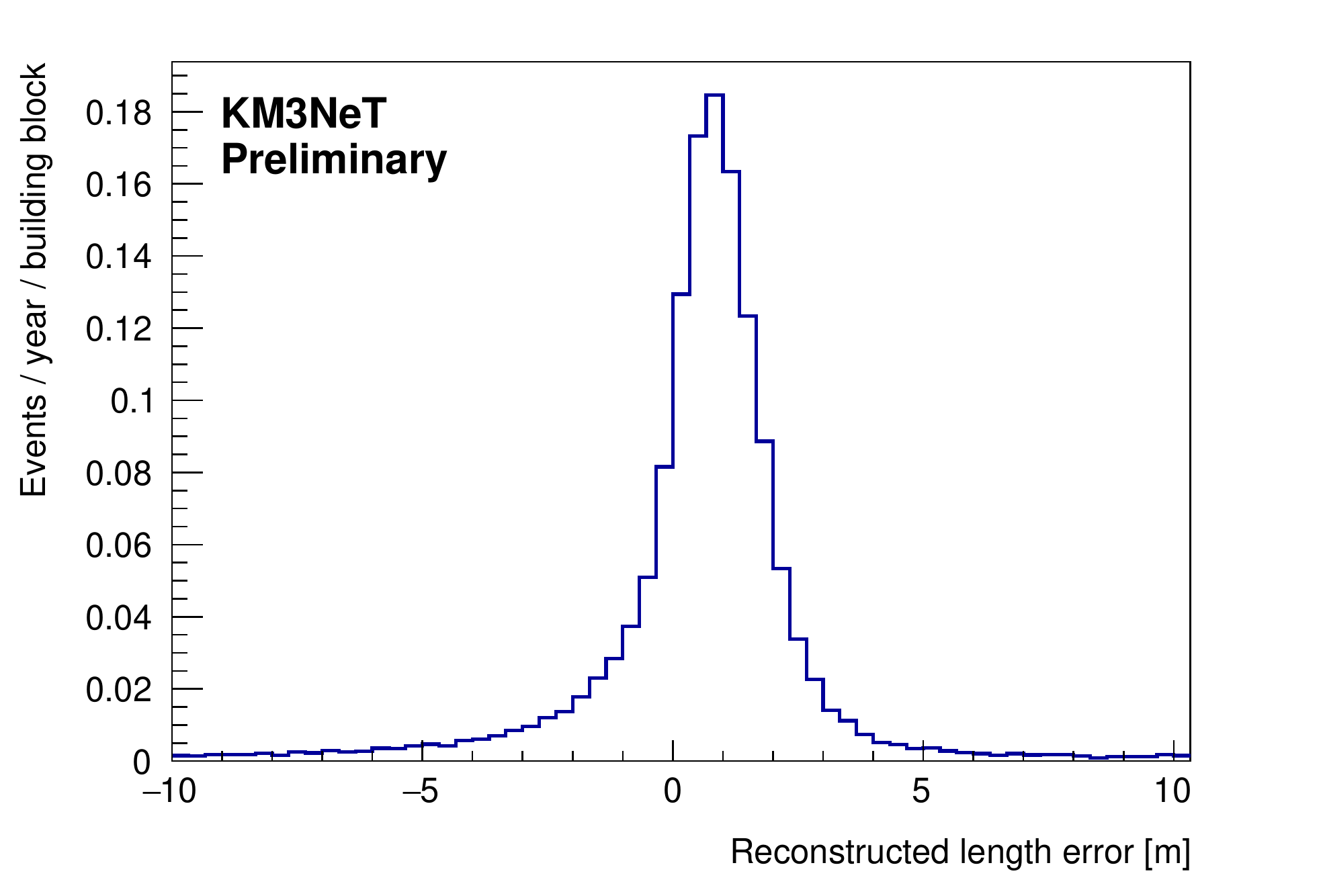}}}%
    \qquad
    \subfloat[\centering Reconstructed energy error for the visible energy. The median and 68\% quantiles are $-1.75^{+6.11}_{-6.90}$\%.]{{\includegraphics[width=7cm]{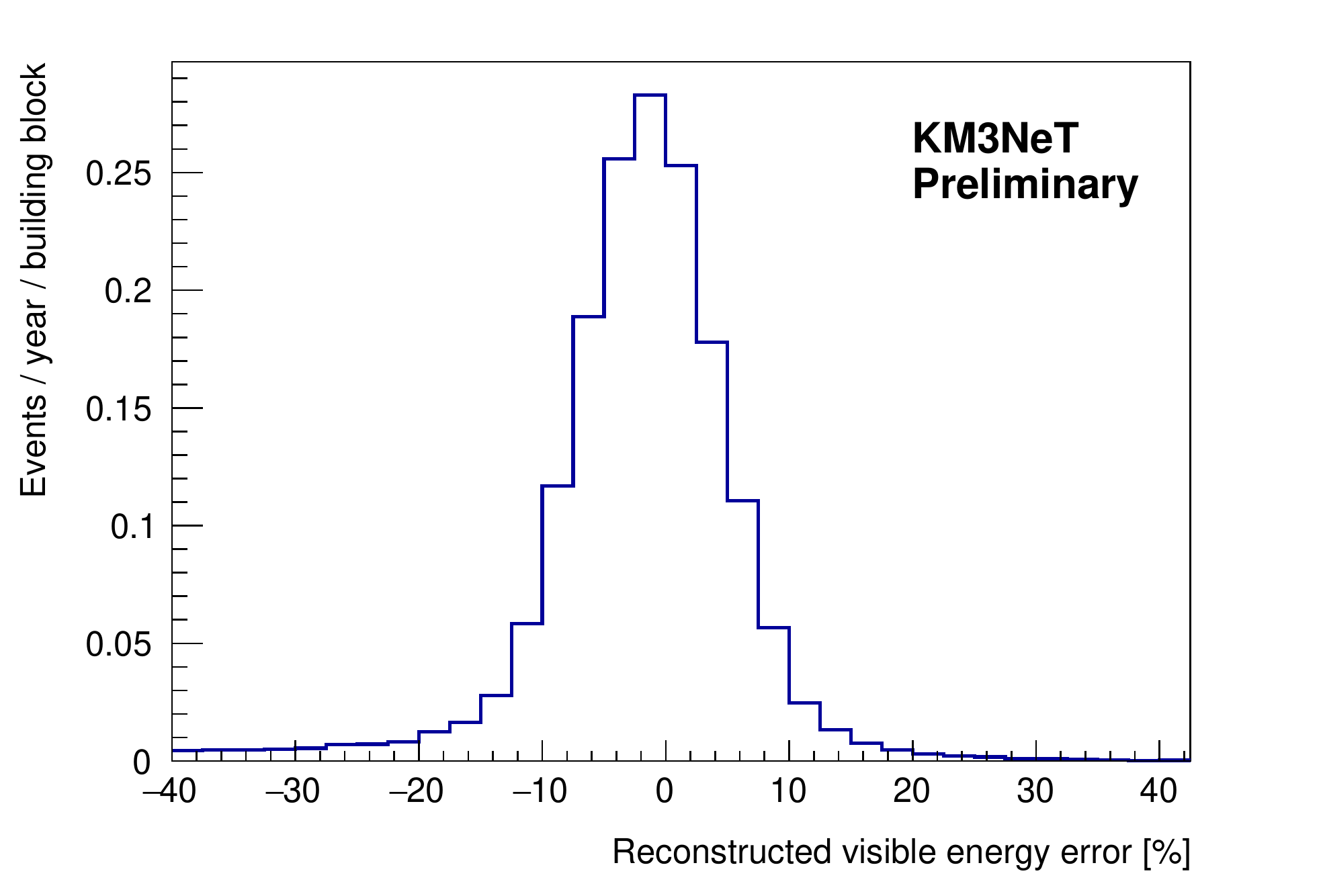}}}%
    \caption{Reconstruction performance for double cascade reconstruction on selected double cascade events.}
    \label{fig:lenreseres}%
\end{figure}

Figure \ref{fig:lenreseres} shows the reconstructed length error and the reconstructed energy error for the double cascade reconstruction. The median and 68\% quantiles for the reconstructed length error are $0.72^{+1.23}_{-1.95}$ meters showing a small bias for overestimating the true tau length. The median and 68\% quantiles for the reconstructed energy error are $-1.75^{+6.11}_{-6.90}$\%.

\section{Summary}

KM3NeT/ARCA shows great potential for the reconstruction of double cascade events. The lever-arm effect improves the angular reconstruction performance to sub-degree level for tau lengths larger than 25 meters. The tau length resolution is 3.17 meters and the energy resolution is 13\%. Event selection is not yet covered in the work and will be the topic of future efforts. This includes the response of the algorithm on atmospheric muon background events and single cascades from neutral current and $\nu_e$ charged current interactions.

\clearpage
\section*{Full Author List: KM3NeT Collaboration}

\scriptsize
\noindent
M.~Ageron$^{1}$,
S.~Aiello$^{2}$,
A.~Albert$^{3,55}$,
M.~Alshamsi$^{4}$,
S. Alves Garre$^{5}$,
Z.~Aly$^{1}$,
A. Ambrosone$^{6,7}$,
F.~Ameli$^{8}$,
M.~Andre$^{9}$,
G.~Androulakis$^{10}$
% ...
\\

\noindent
$^{1}$Aix~Marseille~Univ,~CNRS/IN2P3,~CPPM,~Marseille,~France. \\
$^{2}$INFN, Sezione di Catania, Via Santa Sofia 64, Catania, 95123 Italy. \\
$^{3}$Universit{\'e}~de~Strasbourg,~CNRS,~IPHC~UMR~7178,~F-67000~Strasbourg,~France. \\
$^{4}$Universit{\'e} de Paris, CNRS, Astroparticule et Cosmologie, F-75013 Paris, France. \\
$^{5}$IFIC - Instituto de F{\'\i}sica Corpuscular (CSIC - Universitat de Val{\`e}ncia), c/Catedr{\'a}tico Jos{\'e} Beltr{\'a}n, 2, 46980 Paterna, Valencia, Spain. \\
$^{6}$INFN, Sezione di Napoli, Complesso Universitario di Monte S. Angelo, Via Cintia ed. G, Napoli, 80126 Italy. \\
$^{7}$Universit{\`a} di Napoli ``Federico II'', Dip. Scienze Fisiche ``E. Pancini'', Complesso Universitario di Monte S. Angelo, Via Cintia ed. G, Napoli, 80126 Italy. \\
$^{8}$INFN, Sezione di Roma, Piazzale Aldo Moro 2, Roma, 00185 Italy. \\
$^{9}$Universitat Polit{\`e}cnica de Catalunya, Laboratori d'Aplicacions Bioac{\'u}stiques, Centre Tecnol{\`o}gic de Vilanova i la Geltr{\'u}, Avda. Rambla Exposici{\'o}, s/n, Vilanova i la Geltr{\'u}, 08800 Spain. \\
$^{10}$NCSR Demokritos, Institute of Nuclear and Particle Physics, Ag. Paraskevi Attikis, Athens, 15310 Greece. \\
% ...
$^{55}$Universit{\'e} de Haute Alsace, 68100 Mulhouse Cedex, France. \\
% ...

\end{document}